# ON THE STRUCTURE OF THE EULER MAPPING

Demeter Krupka, Brno

**1. Introduction** Let $\mathbf{R}^n$, $\mathbf{R}^m$ be real Euclidean spaces of dimension $n$, $m$, respectively, $L(\mathbf{R}^n, \mathbf{R}^m)$ the vector space of all linear mappings from $\mathbf{R}^n$ into $\mathbf{R}^m$, $L_s^2(\mathbf{R}^n, \mathbf{R}^m)$ the vector space of all symmetric bilinear mappings from $\mathbf{R}^n$ into $\mathbf{R}^m$, $U \subset \mathbf{R}^n$ and $V \subset \mathbf{R}^m$ some open sets. We write $\mathbf{R} = \mathbf{R}^1$. Put

$$\mathcal{J}^1 = U \times V \times L(\mathbf{R}^n, \mathbf{R}^m), \quad \mathcal{J}^2 = U \times V \times L(\mathbf{R}^n, \mathbf{R}^m) \times L_s^2(\mathbf{R}^n, \mathbf{R}^m)$$

(the cartesian products), and consider $\mathcal{J}^1$ and $\mathcal{J}^2$ as differentiable manifolds with natural coordinates $(x_i, y_\mu, z_{i\mu})$ and $(x_i, y_\mu, z_{i\mu}, z_{ki\mu})$ respectively ($1 \le k \le i \le n$, $1 \le \mu \le m$). Denote by $\Gamma$ the set of all differentiable maps $f : U \to V$ (say, of class $C^2$), and write $D^r f$ for the $r$-th derivative of the map $f$ [2], $r$=1,2.

Assume that we have a real function $L$ on $\mathcal{J}^1$ and a compact domain $\Omega \subset U$. The data give rise to the real function

$$\Gamma \ni f \to \int_\Omega L(x, f(x), Df(x))dx \in \mathbf{R} \tag{1}$$

(with $dx = dx_1 \wedge \ldots \wedge dx_n$) which is of principal interest in various problems of the calculus of variations (see e.g. [3]). The *extremals* associated with $L$ are then defined as solutions $f \in \Gamma$ of the so called *Euler equations*

$$\mathcal{E}_\mu(L) = \frac{\partial L}{\partial y_\mu} - \frac{\partial^2 L}{\partial x_k \partial z_{k\mu}} - \frac{\partial^2 L}{\partial y_\sigma \partial z_{k\mu}} z_{k\sigma} - \frac{\partial^2 L}{\partial z_{i\sigma} \partial z_{k\mu}} z_{ki\sigma} = 0, \tag{2}$$
$$\mu = 1, 2, \ldots, m.$$

Here, is everywhere in this paper, the usual summation convention is used.

The expressions $\mathscr{E}_\mu(L)$ defined by (2) are functions on $\mathscr{J}^2$. Put

$$\omega_\mu = dy_\mu - z_{j\mu}dx_j,$$

and define a 1-form $\mathscr{E}(L)$ on $\mathscr{J}^2$ by the formula

$$\mathscr{E}(L) = \mathscr{E}_\mu(L)\omega_\mu .$$

It can be easily checked that the 1-form $\mathscr{E}(L)$ is independent of the particular choice of coordinates on $\mathscr{J}^2$.

We shall call each function $L$ on $\mathscr{J}^1$ the *Lagrange function*, and the 1-form $\mathscr{E}(L)$ the *Euler form* associated with the Lagrange function $L$. The vector space (over $\mathbf{R}$) of all Lagrange functions is denoted by $\mathscr{L}(\mathscr{J}^1)$, and the vector space of all 1-forms on $\mathscr{J}^2$ (over $\mathbf{R}$) is denoted by $\Omega^1(\mathscr{J}^2)$.

Certain sufficient conditions for the identical vanishing of the left-hand sides of the Euler equations (2), or, which is the same, for $\mathscr{E}(L) = 0$, are known and frequently used in various calculations. Suppose that $L \in \mathscr{L}(\mathscr{J}^1)$ is of the form of the so called "divergence expression"

$$L = \frac{\partial f_i}{\partial x_i} + \frac{\partial f_i}{\partial x_\sigma} z_{i\sigma}, \tag{3}$$

where $f_i$, $1 \le i \le n$, are some functions on $U \times V$. Then we see at once that $\mathscr{E}(L) = 0$. It is also known that for the case $m = 1$ condition (3) is necessary: this is a classical proposition of Courant and Hilbert [1].

We mention just two cases when condition (3) is used:

1. In the classical mechanics [5] and the general relativity [6], (3) serves for replacing the given Lagrange function by a more simple one.

2. In the theory of invariant variational problems, for definition of the so called generalized invariant transformations [8] (see also [4], [9]).

On the other hand, a complete description of the Lagrange functions $L$ satisfying $\mathscr{E}(L) = 0$ has not yet been given unless $m = 1$. The goal of this paper is to give such a description. In other words we shall study the kernel of the linear mapping

$$\mathscr{L}(\mathscr{J}^1) \ni L \to \mathscr{E}(L) \in \Omega^1(\mathscr{J}^2)$$

which will be referred to as the *Euler mapping*.

**2. Definitions and lemmas.** For the purpose of this paper it suffices to define what we mean by horizontal differential forms on the cartesian product of open subsets of Euclidean spaces.

Let $V$ and $W$ be some open sets in the finite dimensional Euclidean spaces $\mathbf{R}^n$ and $\mathbf{R}^m$, and consider the cartesian product $V \times W$ and the natural projection $\pi: V \times W \to V$ on the first factor. A tangent vector $\xi$ at a point $(v, w) \in V \times W$ is called $\pi$-*vertical*, if

$$D\pi(v, w) \cdot \xi = 0.$$

A differential form $\rho$ on $V\times W$ is called $\pi$-*horizontal* if it vanishes whenever one of its arguments (i.e. tangent vectors) is a $\pi$-vertical vector.

Let us turn to the notation of Introduction.

We denote $\pi_1:\mathcal{J}^1\to U$, $\pi_{10}:\mathcal{J}^1\to U\times V$, $\pi_{20}:\mathcal{J}^2\to U\times V$ the natural projections and shall therefore speak about $\pi_1$-horizontal, $\pi_{10}$-horizontal, and $\pi_{20}$-horizontal differential forms. Correspondingly, we shall write $\Omega^n_U(\mathcal{J}^1)$, $\Omega^n_{U\times V}(\mathcal{J}^1)$, and $\Omega^1_{U\times V}(\mathcal{J}^2)$ for the spaces of all $\pi_1$-horizontal $n$-forms, $\pi_{10}$-horizontal $n$-forms, and $\pi_{20}$-horizontal 1-forms (remember that $n=\dim U$).

Notice that the Euler form , $\mathcal{E}(L)$, is an element of $\Omega^1_{U\times V}(\mathcal{J}^2)$.

Let $f\in\Gamma$ define the mapping

$$U\ni x\to jf(x)=(x,f(x),Df(x))\in\mathcal{J}^1$$

and denote by $jf^*$ the corresponding mapping induced on differential forms on $\mathcal{J}^1$. Thus, if $\rho$ is a differential $p$-form on $\mathcal{J}^1$, then $jf^*\rho$ is a differential $p$-form on $U$.

**Lemma 1.** *There is one and only one mapping*

$$\Omega^n_{U\times V}(\mathcal{J}^1)\ni\rho\to h(\rho)\in\Omega^n_U(\mathcal{J}^1)$$

*satisfying the following two conditions:*

1. *$h$ is linear over the ring of functions on $\mathcal{J}^1$*
2. *If $\rho\in\Omega^n_{U\times V}(\mathcal{J}^1)$ is an arbitrary n-form, then*

$$jf^*\rho = jf^*h(\rho)$$

*for all $f\in\Gamma$.*

**Proof.** If the mapping $h$ exists, it is obviously unique. Let $\rho$ be an arbitrary element of $\Omega^n_{U\times V}(\mathcal{J}^1)$. If in the natural coordinates $(x_i,y_\mu,z_{i\mu})$, $\rho$ has the expression

$$\rho = g_0dx_1\wedge\ldots\wedge dx_n+\sum_{r=1}^n\sum_{s_1<\ldots<s_r}\sum_{\sigma_1,\ldots,\sigma_r}\frac{1}{r!}g^{s_1}_{\sigma_1}\cdots^{s_r}_{\sigma_r}\,dx_1\wedge\ldots\wedge dx_{s_1-1}\wedge dy_{\sigma_1}\wedge dx_{s_1+1}\wedge\ldots\wedge dy_{\sigma_r}\wedge\ldots\wedge dx_n \tag{4}$$

(in which the function $g^{s_1}_{\sigma_1}\cdots^{s_r}_{\sigma_r}$ are supposed to be antisymmetric in all subscripts), then we define

$$h(\rho)=\left(g_0+\sum_{r=1}^n\sum_{s_1<\ldots<s_r}\sum_{\sigma_1,\ldots,\sigma_r}g^{s_1}_{\sigma_1}\cdots^{s_r}_{\sigma_r}\,z_{s_1\sigma_1}\ldots z_{s_r\sigma_r}\right)dx_1\wedge\ldots\wedge dx_n \tag{5}$$

It is immediately clear that conditions 1 and 2 are satisfied.

**Lemma 2.** *The mapping $h$ is surjective.*

**Proof.** Let

$$\rho=Ldx_1\wedge\ldots\wedge dx_n \tag{6}$$

be an arbitrary $\pi_1$-horizontal $n$-form. We take

$$\lambda = L dx_1 \wedge \ldots \wedge dx_n + \frac{\partial L}{\partial z_{i\sigma}} \omega_\sigma^i ,$$

where

$$\omega_\sigma^i = dx_1 \wedge \ldots \wedge dx_{i-1} \wedge (dy_\sigma - z_{k\sigma} dx_k) \wedge dx_{i+1} \wedge \ldots \wedge dx_n .$$

The equality $h(\rho) = \lambda$ follows from (5).

We note that the form $\rho$ from the proof is invariant under coordinate transformations on $\mathcal{J}^1$. It has been introduced, in a special case, by Sniatycki [7] in connection with some geometric considerations concerning the structure of the calculus of variations.

In order to shorten the proof of our main theorem we state the following explicit formula for the exterior differential $d\rho$ of a form $\rho \in \Omega^n_{U\times V}(\mathcal{J}^1)$.

**Lemma 3.** *Let* $\rho \in \Omega^n_{U\times V}(\mathcal{J}^1)$ *be expressed as in* (4). *Then* $d\rho$ *is expressed as*

$$\begin{aligned}
d\rho = {} & \left( \frac{\partial g_0}{\partial y_\mu} - \frac{\partial g^s_\mu}{\partial x_s} \right) dy_\mu \wedge dx_1 \wedge \ldots \wedge dx_n + \sum_{r=1}^{n-1} \sum_{s_1 < \ldots < s_r} \sum_{\sigma_1, \ldots, \sigma_r} \frac{1}{(r+1)!} \\
& \cdot \left( \frac{\partial g^{s_1 \ldots s_r}_{\sigma_1 \ldots \sigma_r}}{\partial y_\mu} - \frac{\partial g^{s_1 s_2 \ldots s_r}_{\mu \sigma_2 \ldots \sigma_r}}{\partial y_{\sigma_1}} - \ldots - \frac{\partial g^{s_1 \ldots s_r}_{\sigma_1 \ldots \sigma_r}}{\partial y_{\sigma_r}} \sum_{k + s_1} \frac{\partial g^{k s_1 \ldots s_r}_{\mu \sigma_1 \ldots \sigma_r}}{\partial x_k} \right. \\
& \left. - \sum_{s_1 < k < s_2} \frac{\partial g^{s_1 k s_2 \ldots s_r}_{\sigma_1 \mu \sigma_2 \ldots \sigma_r}}{\partial x_k} - \ldots - \sum \frac{\partial g^{s_1 \ldots s_r k}_{\sigma_1 \ldots \sigma_r \mu}}{\partial x_k} \right) dy_\mu \wedge dx_1 \wedge \ldots \wedge dx_{s_1 - 1} \\
& \wedge dy_{\sigma_1} \wedge dx_{s_1 + 1} \wedge \ldots \wedge dx_{\sigma_r} \wedge \ldots \wedge dx_n \\
& + \sum_{\sigma_1, \ldots, \sigma_n} \frac{1}{(n+1)!} \left( \frac{\partial g^{1 \ldots n}_{\sigma_1 \ldots \sigma_n}}{\partial y_\mu} - \frac{\partial g^{12 \ldots n}_{\mu \sigma_2 \ldots \sigma_n}}{\partial y_{\sigma_1}} - \ldots - \frac{\partial g^{1 \ldots n}_{\sigma_1 \ldots \mu}}{\partial y_{\sigma_n}} \right) \\
& \cdot dy_\mu \wedge dy_{\sigma_1} \wedge \ldots \wedge dy_{\sigma_n} + \frac{\partial g_0}{\partial z_{k\mu}} dz_{k\mu} \wedge dx_1 \wedge \ldots \wedge dx_n \\
& + \sum_{r=1}^{n} \sum_{s_1 < \ldots < s_r} \sum_{\sigma_1, \ldots, \sigma_r} \frac{1}{r!} \frac{\partial g^{s_1 \ldots s_r}_{\sigma_1 \ldots \sigma_r}}{\partial z_{k\mu}} dz_{k\mu} \\
& \wedge dx_1 \wedge \ldots \wedge dx_{s_1 - 1} \wedge dy_{\sigma_1} \wedge dx_{s_1 + 1} \wedge \ldots \wedge dx_{\sigma_r} \wedge \ldots \wedge dx_n .
\end{aligned} \tag{7}$$

**Proof.** The formula follows by a straightforward calculation.

**3. The kernel of the Euler mapping.** The main result of this work is contained in the following:

**Theorem.** *Let* $L \in \mathcal{L}(\mathcal{J}^1)$ *be a Lagrange function. Then the following two conditions are equivalent:*

1. *The Euler form associated with* $L$ *vanishes,* $\mathcal{E}(L) = 0$.
2. *There exists an* $n$*-form* $\rho \in \Omega^n_{U\times V}(\mathcal{J}^1)$ *such that*

a) $h(\rho) = L dx_1 \wedge \ldots \wedge dx_n$,

b) $d\rho = 0$.

*The n-form $\rho$ is uniquely determined by $L$.*

**Proof.** Suppose that $\mathscr{E}(L) = 0$. Then the relations (2) hold for all $(x_i, y_\mu, z_{i\mu}, z_{ki\mu})$, and are equivalent with the system

$$(8) \qquad \frac{\partial^2 L}{\partial z_{i\sigma}\,\partial z_{k\mu}} + \frac{\partial^2 L}{\partial z_{k\sigma}\,\partial z_{i\mu}} = 0,$$

$$(9) \qquad \frac{\partial L}{\partial y_\mu} - \frac{\partial^2 L}{\partial x_k\,\partial z_{k\mu}} - \frac{\partial^2 L}{\partial y_\sigma\,\partial z_{k\mu}} z_{k\sigma} = 0.$$

From the first condition (8) we find that $L$ must be of the form

$$(10) \qquad L = f_0 + \sum_{r=1}^{n} \sum_{s_1<\ldots<s_r} \sum_{\sigma_1,\ldots,\sigma_r} f^{s_1 \cdots s_r}_{\sigma_1 \cdots \sigma_r} \cdot z_{s_1\sigma_1} \ldots z_{s_r\sigma_r},$$

where $f_0$ and $f^{s_1 \cdots s_r}_{\sigma_1 \cdots \sigma_r}$ do not depend on $z_{j\mu}$ and $f^{s_1 \cdots s_r}_{\sigma_1 \cdots \sigma_r}$ are antisymmetric in $\sigma_1, \ldots, \sigma_r$. Let us examine the second condition (9). After some calculation we get

$$(11) \qquad \begin{aligned} &\frac{\partial f_0}{\partial y_\mu} - \frac{\partial f^k_\mu}{\partial x_k} + \sum_{r=1}^{n-1} \sum_{s_1<\ldots<s_r} \sum_{\sigma_1,\ldots,\sigma_r} \left( \frac{\partial f^{s_1 \cdots s_r}_{\sigma_1 \cdots \sigma_r}}{\partial y_\mu} - \frac{\partial f^{s_1 s_2 \cdots s_r}_{\mu\sigma_2 \cdots \sigma_r}}{\partial y_{\sigma_1}} - \ldots \right. \\ &- \frac{\partial f^{s_1 \cdots s_r}_{\sigma_1 \cdots \mu}}{\partial y_{\sigma_r}} - \sum_{k<s_1} \frac{\partial f^{k s_1 \cdots s_r}_{\mu\sigma_1 \cdots \sigma_r}}{\partial x_k} - \sum_{s_1<k<s_2} \frac{\partial f^{s_1 k s_2 \cdots s_r}_{\sigma_1\mu\sigma_2 \cdots \sigma_r}}{\partial x_k} - \ldots \\ &\left. - \sum_{k>s_r} \frac{\partial f^{s_1 \cdots s_r k}_{\sigma_1 \cdots \sigma_r \mu}}{\partial x_k} \right) z_{s_1\sigma_1} \ldots z_{s_r\sigma_r} + \sum_{\sigma_1,\ldots,\sigma_n} \left( \frac{\partial f^{1 \cdots n}_{\sigma_1 \cdots \sigma_n}}{\partial y_\mu} - \right. \\ &\left. \frac{\partial f^{12 \cdots n}_{\mu\sigma_2 \cdots \sigma_n}}{\partial y_{\sigma_1}} - \ldots - \frac{\partial f^{1 \cdots n}_{\sigma_1 \cdots \mu}}{\partial y_{\sigma_n}} \right) z_{1\sigma_1} \ldots z_{n\sigma_n} = 0. \end{aligned}$$

Since the coefficients at $z_{s_1\sigma_1} \ldots z_{s_r\sigma_r}$ do not depend on $z_{k\mu}$ they must vanish separately. In this way we have obtained that if $L$ satisfies $\mathscr{E}(L) = 0$, then $L$ is of the form (10) and the conditions (11) are satisfied. We assert that the functions $f_0$, $f^{s_1 \cdots s_r}_{\sigma_1 \cdots \sigma_r}$ are unique: it follows from (10) that

$$f^{1 \cdots n}_{\nu_1 \cdots \nu_n} = \frac{\partial^n L}{\partial z_{1\nu_1} \ldots \partial z_{n\nu_n}},$$

$$\ldots$$

$$f^{s_1 \cdots s_r}_{\nu_1 \cdots \nu_r} = \frac{\partial^r L}{\partial z_{s_1\nu_1} \ldots \partial z_{s_r\nu_r}} - \sum_{j=r+1}^{\nu} \sum_{k_1<\ldots<k_j} \sum_{\sigma_1,\ldots,\sigma_j} f^{k_1 \cdots k_j}_{\sigma_1 \cdots \sigma_j}$$

$$\cdot \frac{\partial^j}{\partial z_{s_1\nu_1} \ldots \partial z_{s_r\nu_r}} (z_{k_1\sigma_1} \ldots z_{k_j\sigma_j}),$$

$$\ldots$$

$$f_0 = L - \sum_{r=1}^{n} \sum_{s_1<\ldots<s_r} \sum_{\sigma_1,\ldots,\sigma_r} f_{\sigma_1}^{s_1} \cdots_{\sigma_r}^{s_r} z_{s_1\sigma_1} \ldots z_{s_r\sigma_r}.$$

Consequently, if we put

$$\rho = f_0 dx_1 \wedge \ldots \wedge dx_n + \sum_{r=1}^{n} \sum_{s_1<\ldots<s_r} \sum_{\sigma_1,\ldots,\sigma_r} \frac{1}{r!} f_{\sigma_1}^{s_1} \cdots_{\sigma_r}^{s_r}$$
$$\cdot\, dx_1 \wedge \ldots \wedge dx_{s_1-1} \wedge dy_{\sigma_1} \wedge dx_{s_1+1} \wedge \ldots \wedge dy_{\sigma_r} \wedge \ldots \wedge dx_n$$

we obtain, by (5) and Lemma 3, that condition 2 from the theorem is satisfied by $\rho$. At the same time we have proved that the $n$-form $\rho$ is unique.

Conversely, suppose that we have an $n$-form $\rho \in \Omega^n_{U\times V}(\mathscr{J}^1)$ satisfying 2. By comparison with Lemma 3 it can be seen at once that the Lagrange function $L$ defined by 2 a) satisfies the condition 1.

This proves the Theorem.

**Remark 1.** Let $L \in \mathscr{L}(\mathscr{J}^1)$ be a Lagrange function satisfying the condition $\mathscr{E}(L) = 0$, and $\rho$ the corresponding $n$-form from the Theorem. Since the functions $f_0$, $f_{\sigma_1}^{s_1} \cdots_{\sigma_r}^{s_r}$ do not depend on $z_{i\mu}$, the form $\rho$ can be regarded as defined on $U \times V$. The property $d\rho = 0$ then means that we can find, at least locally, an $(n-1)$-form $\eta$ on $U \times V$ such that

$$\rho = d\eta.$$

(This follows from the well-known Poincare lemma concerning the so called closed forms.) We thus observe that $L$ satisfies the relation

(12) $$L dx_1 \wedge \ldots \wedge dx_n = h(d\eta).$$

Conversely, if we take an arbitrary $(n-1)$-form $\eta$ defined on $U \times V$ and *define* $L$ by relation (12) we can see at once that the function $L$ leads to the equality $\mathscr{E}(L) = 0$.

Thus, having in mind Lemma 2, we can say that condition (12) with arbitrary $(n-1)$-forms $\eta$ on $U \times V$ describes all the Lagrange functions for which $\mathscr{E}(L) = 0$.

**Remark 2.** We note that all considerations from this paper can be extended to the case when there is given a fibred manifolds $(Y, \pi, X)$, and Lagrange functions defined on the first jet prolongation of the fibred manifold are considered (see [4] and [8]).

**Remark 3.** a) If $n = 1$, then $0$-forms on $U \times V$ are just real functions. If we write $(x, y_\mu, \dot{y}_\mu)$ for the natural coordinates on $\mathscr{J}^1$ in this case we get, for an arbitrary function $F$ on $U \times V$,

$$h(dF) = \left( \frac{\partial F}{\partial x} + \frac{\partial F}{\partial y_\mu} \dot{y}_\mu \right) dx,$$

and

$$L = \frac{\partial F}{\partial x} + \frac{\partial F}{\partial y_\mu} y\!\!\!x_\mu .$$

b) If $n = 2$, then the general 1-form on $U \times V$ can be expressed as

$$\eta = f_i dx_i + g_\mu dy_\mu .$$

After some calculation

$$h(d\eta) = \left( \frac{\partial f_j}{\partial x_i} + \left( \frac{\partial g_\mu}{\partial x_i} - \frac{\partial f_i}{\partial y_\mu} \right) z_{j\mu} + \frac{\partial g_\mu}{\partial y_\sigma} z_{i\sigma} z_{j\mu} \right) \varepsilon_{ij} \cdot dx_1 \wedge dx_2 .$$

In this formula $\varepsilon_{11} = \varepsilon_{22} = 0$, $\varepsilon_{12} = -\varepsilon_{21} = 1$. The Lagrange functions $L$ leading to zero Euler form have therefore to be of the form

$$L = \left( \frac{\partial f_j}{\partial x_i} + \left( \frac{\partial g_\mu}{\partial x_i} - \frac{\partial f_i}{\partial y_\mu} \right) z_{j\mu} + \frac{\partial g_\mu}{\partial y_\sigma} z_{i\sigma} z_{j\mu} \right) \varepsilon_{ij} .$$

c) If $n$ is general, one can proceed in the same manner as in the case a) or b).

*D. Krupka*
*611 37 Brno, Kotlarska 2*
*Czechoslovakia*